# Secure Multicast Key Distribution for Mobile Adhoc Networks

D.SuganyaDevi
Asst.Prof, Department of Computer Applications
SNR SONS College
Coimbatore, Tamil Nadu, India
.

Dr.G.Padmavathi
Prof. and Head, Dept. of Computer Science,
Avinashilingam University for Women,
Coimbatore, Tamil Nadu, India
.

*Abstract*— **Many emerging applications in mobile adhoc networks involve group-oriented communication. Multicast is an efficient way of supporting group oriented applications, mainly in mobile environment with limited bandwidth and limited power. For using such applications in an adversarial environment as military, it is necessary to provide secure multicast communication. Key management is the fundamental challenge in designing secure multicast communications. In many multicast interactions, new member can join and current members can leave at any time and existing members must communicate securely using multicast key distribution within constrained energy for mobile adhoc networks. This has to overcome the challenging element of "1 affects n" problem which is due to high dynamicity of groups. Thus this paper shows the specific challenges towards multicast key management protocols for securing multicast key distribution in mobile ad hoc networks, and present relevant multicast key management protocols in mobile ad hoc networks. A comparison is done against some pertinent performance criteria.**

## I. INTRODUCTION

A MANET (Mobile Adhoc Network) is an autonomous collection of mobile users that offers infrastructure-free communication over a shared wireless medium. It is formed spontaneously without any preplanning. Multicasting is a fundamental communication paradigm for group-oriented communications such as video conferencing, discussion forums, frequent stock updates, video on demand (VoD), pay per view programs, and advertising.

The combination of an adhoc environment [1, 2] with multicast services induces new challenges towards the security infrastructure to enable acceptance and wide deployment of multicast communication. Indeed, several sensitive applications based on multicast communications have to be secured within adhoc environments. For example military applications such as group communication in a battlefield and also public security operations involving fire brigades and policemen have to be secured.

To prevent attacks and eavesdropping, basic security services such as authentication, data integrity, and group confidentiality are necessary for collaborative applications. Among which group confidentiality is the most important service for military applications. These security services can be facilitated if group members share a common secret, which in turn makes key management [3] a fundamental challenge in designing secure multicast communication systems.

To ensure group confidentiality during the multicast session, the sender (source) shares a secret symmetric key with all valid group members, called Traffic Encryption Key (TEK). To multicast a secret message, the source encrypts the message with the TEK using a symmetric encryption algorithm. Upon receiving the encrypted multicast message, each valid member that knows the TEK can decrypt it with TEK and recover the original one. Key management includes creating, distributing and updating the keys then it constitutes a basic block for secure multicast communication applications.

Each member holds a key to encrypt and decrypt the multicast data. When a member joins and leaves a group, the key has to be updated and distributed to all group members in order to meet the above requirements. The process of updating the keys and distributing them to the group members is called rekeying operation [4]. Rekeying is required in secure multicast to ensure that a new member cannot decrypt the stored multicast data (before its joining) and prevents a leaving member from eavesdropping future multicast data.

A critical problem with any rekey technique is scalability. The rekey process should be done after each membership change, and if the membership changes are frequent, key management will require a large number of key exchanges per unit time in order to maintain both forward and backward secrecies. The number of TEK update messages in the case of frequent join and leave operations induces "1 affects n" phenomenon [5].

To overcome this problem, several approaches propose a multicast group clustering [5,6 and 7]. Clustering is dividing the multicast group into several sub-groups. A Local Controller (LC) manages each sub group, which is responsible for local key management within the cluster. Thus, after Join or Leave procedures, only members within the concerned cluster are affected by rekeying process, and the local dynamics of a cluster does not affect the other clusters of the group. Moreover, few solutions for multicast group clustering did consider the energy and latency issues to achieve an efficient key distribution process, whereas energy and latency constitutes main issue in ad hoc environments. This paper extends and presents taxonomy of multicast key distribution protocols, dedicated to operate in ad hoc networks for secure multicast communications.

The remainder of this paper is structured as follows. Section 2 emphasizes the challenges of securing multicast communications within ad hoc environments. Section 3 presents the key management requirements. Section 4 describes Taxonomy of Multicast key management approaches. Section 5 discusses the approaches. Finally, Section 6 concludes the paper.

## II. CHALLENGES AND CONSTRAINTS OF SECURING MULTICAST KEY DISTRIBUTION FOR MOBILE AD HOC NETWORKS

The principal constraints and challenges induced by the ad hoc environment [8] are as follows.

- **Wireless Links**: The wireless links make the network easily prone to passive malicious attacks like sniffing, or active attacks like message replay or message alteration.

- **Absence of Infrastructure**: The absence of infrastructure is one of the main characteristics of ad hoc networks.

- **Autonomous** No centralized administration entity is available to manage the operation of the different mobile nodes.

- **Dynamic topology** Nodes are mobile and can be connected dynamically in an arbitrary manner. Links of the network vary timely and are based on the proximity of one node to another node.

- **Device discovery** Identifying relevant newly moved in nodes and informing about their existence need dynamic update to facilitate automatic optimal route selection.

- **Bandwidth optimization** Wireless links have significantly lower capacity than the wired links.

- **Limited Power**: Adhoc networks are composed of low powered devices. These devices have limited energy, bandwidth and CPU, as well as low memory capacities.

- **Scalability** defined as whether the network is able to provide an acceptable level of service even in the presence of a large number of nodes.

- **Self operated** Self healing feature demands MANET should realign itself to blanket any node moving out of its range.

- **Poor Transmission Quality** This is an inherent problem of wireless communication caused by several error sources that result in degradation of the received signal.

- **Ad hoc addressing** Challenges in standard addressing scheme to be implemented.

- **Network configuration** The whole MANET infrastructure is dynamic and is the reason for dynamic connection and disconnection of the variable links.

- **Topology maintenance** Updating information of dynamic links among nodes in MANETs is a major challenge.

Consequently, achieving secure multicast communications in adhoc networks should take into account additional factors including the energy consumption efficiency, the optimal selection of group controllers and saves the bandwidth.

## III. KEY MANAGEMENT REQUIREMENTS

Key management includes creating, distributing and updating the keys then it constitutes a basic block for secure multicast communication applications. Group confidentiality requires that only valid users could decrypt the multicast data. Efficient key management protocols should take into consideration of miscellaneous requirements [4]. Figure 1 summarizes these.

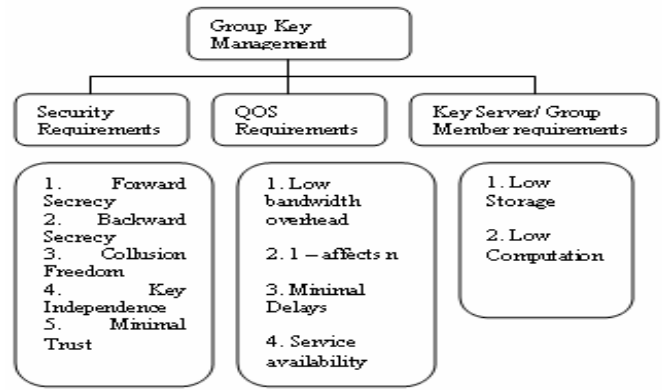

Figure 1. Group Key Management Requirements

### A. Security requirements

- **Forward secrecy** This ensures that a member cannot decrypt data after it leaves the group. To assure forward secrecy, a re-key of the group with a new TEK after each leave from the group is the ultimate solution.

- **Backward secrecy** This ensures that a member cannot decrypt data sent before it joins the group. To assure backward secrecy, a re-key of the group with a new TEK after each join to the group is the ultimate solution.

- **Collusion freedom** requires that any set of fraudulent users should not be able to deduce the current traffic encryption key.

- **Key independence**: This ensures that any subset of a group keys must not be able to discover any other group key.

- **Trust relationship**: In mobile ad hoc groups there is no trusted central authority that is actively involved in the computation of group key that is all participants have equal rights during computation process. This is emphasized by definition of verifiable trust relationship that consists of two requirements: One as Group members are trusted not to reveal the group key or secret values that may lead to its computation to any other party, and another as group members must be able to verify the computation steps of the group key management protocol.

### B. Quality of service requirement

- **Low bandwidth overhead**: the re-key of the group should not induce a high number of messages, especially for dynamic groups. Ideally, this should be independent from the group size.

- **1-affects-n**: a protocol suffers from the 1-affects-n phenomenon if a single membership change in the group affects all the other group members. This happens typically when a single membership change requires that all group members commit to a new TEK.

- **Minimal delays**: many applications that are built over the multicast service (typically, multimedia applications) are sensitive to jitters and delays in packet delivery. Therefore, any key management scheme should take this into consideration and hence minimizes the impact of key management on the delays of packet delivery.

- **Service availability**: the failure of a single entity in the key management architecture must not prevent the operation of the whole multicast session.

### C. Key server and Group Member requirements

The key management scheme induces high storage of keys and high computation overhead at the key server or group members.

Thus securing multicast group communication in ad hoc network should focus on both security and Qos requirements.

## IV. KEY MANAGEMENT APPROACHES

Key management approaches can be classified into three classes: centralized, distributed or decentralized. Figure 2 illustrates this classification.

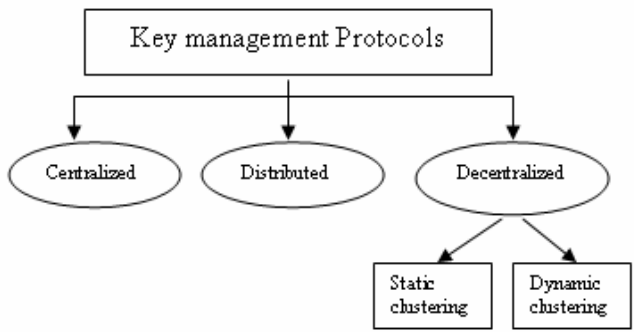

Figure 2. Classification of key management Approaches

### A. Centralized Approaches

In centralized approaches, a designated entity (e.g., the group leader or a key server) is responsible for calculation and distribution of the group key to all the participants. Centralized protocols are further classified into three sub-categories namely Pairwise key approach; Secure locks and Hierarchy of keys approach.

*1.Pairwise key approach*: In this approach, the key server shared pairwise keys with each participant. For example, in GKMP [9], apart from pairwise keys and the group key, all current group participants know a group key encryption key (gKEK). If a new participant joins the group, the server generates a new group key and a new gKEK. These keys are sent to the new member using the key it shares with key server, and to the old group member using the old gKEK.

*2.Secure Locks*: Chiou and Chen [10] proposed Secure Lock; a key management protocol where the key server requires only a single broadcast to establish the group key or to re-key the entire group in case of a leave. This protocol minimizes the number of re-key messages. However, it increases the computation at the server due to the Chinese Remainder calculations before sending each message to the group.

*3. Hierarchy of Keys Approach*: Most efficient approach to rekeying in the centralized case is the hierarchy of keys approach. Here, the key server shares keys with subgroups of the participants, in addition to the pair wise keys. Thus, the hierarchical approach trades off storage for number of transmitted messages.

Logical key hierarchy was proposed independently in [11]. The key server maintains a tree with subgroup keys in the intermediate nodes and the individual keys in the leaves. Apart from the individual keys shared with the key server, each node knows all keys on the path to the root. In root, the group key is stored. As the depth of the balanced binary tree is logarithmical in the number of the leaves, each member stores a logarithmical number of keys, and the number of rekey messages is also logarithmic in the number of group members instead of linear, as in previously described approaches.

One-way function trees (OFT) [12] enables the group members to calculate the new keys based on the previous keys using a one-way function, which further reduces the number of rekey messages.

TABLE I. CENTRALIZED APPROACHES

| Category | Protocol | 1-affects-n | Rekey Overhead | Storage Overhead | | Results |
|---|---|---|---|---|---|---|
| | | | | Key server | Member | |
| Pairwise Keys | GKMP | Yes | 2 | $n+2$ | 3 | Rekey overhead |
| Secrets | Secure Lock | No | 2 | $2n$ | 2 | Computation overhead |
| Hierarchy of keys | LKH | Yes | $Log_2(n)-1$ | $2n-1$ | $Log_2(n)+1$ | Storage complexity |
| | OFT | Yes | $Log_2(n)+1$ | $2n-1$ | $Log_2(n)+1$ | Storage complexity |

In table 1, the pair wise key approach exhibits linear complexity. Secure lock, although most efficient in number of messages, poses serious load on the server and can be used only for small groups. All tree-based protocols have logarithmic communication and storage complexity at the members, and linear storage complexity at the key server.

### B. Distributed Key-Agreement Approaches

With distributed or contributory key-agreement protocols, the group members cooperate to establish a group key. This improves the reliability of the overall system and reduces the bottlenecks in the network in comparison to the centralized approach. The protocols of this category are classified into three sub-categories namely Ring based cooperation, Hierarchical based cooperation and Broadcast based cooperation depending on the virtual topology created by the members for cooperation.

Table 2 shows the comparison results of Distributed Key-Agreement Approaches.

TABLE II.  DISTRIBUTED KEY-AGREEMENT APPROACHES

| Category | Protocol | Verifiable Trust | Leader Required | Results |
|---|---|---|---|---|
| Ring based | Cliques | No | No | Suitable for Low security Level |
| Hierarchical based | STR | Yes | No | Suitable for High security Level |
| Broadcast based | BD | No | No | Suitable for Low security Level |

*1.Ring-Based Cooperation*: In some protocols, members are organized in a ring. The CLIQUES protocol suite [5] is an example of ring-based cooperation. This protocol arranges group members as $(M_1, M_n)$ and $M_n$ as controller. It specifies a role of the controller that collects contributions of other group members, adds own contribution, and broadcasts information that allows all members to compute the group key. The choice of the controller depends on the dynamic event and the current structure. In additive events new members are appended to the end of the list CLIQUES do not provide verifiable trust relationship, because no other member can check whether values forwarded by $M_i$, or the set broadcasted by the controller are correctly built.

*2.Hierarchical Based Cooperation*: In the hierarchical GKA protocols, the members are organized according to some structure.

STR protocol [13] uses the linear binary tree for cooperation and provides communication efficient protocols with especially efficient join and merges operations. STR defines the role of the sponsor temporarily and it can be assigned to different members on dynamic events depending on the current tree structure. The sponsor reduces the communication overhead as it performed some operations on behalf of the group. The sponsor is not a central authority. STR provides verifiable trust relationship because every broadcasted public key can be verified by at least one other participant.

*3.Broadcast based Cooperation*: Broadcast based protocols have constant number of rounds. For example, in three-round Burmester-Desmedt (BD) protocol [14] each participant broadcasts intermediate values to all other participants in each round. The communication and computational load is shared equally between all parties. This protocol does not provide verifiable trust relationship, since no other group member can verify the correctness of the broadcasted values.

### C. Decentralized Approaches

The decentralized approach divides the multicast group into subgroups or clusters, each sub-group is managed by a LC (Local Controller) responsible for security management of members and its subgroup. Two kinds of decentralized protocols are distinguished as static clustering and dynamic clustering.

Table 3 shows the comparison results of Decentralized Approaches.

TABLE III.  DECENTRALIZED APPROACHES

| Category | Protocol | 1-affects-n | Local Re-key | Results |
|---|---|---|---|---|
| Static Clustering | IOLUS | Yes | Yes | More scalable |
| | DEP | Yes | No | More scalable |
| Dynamic Clustering | AKMP & SAKM | No | Yes | Suitable to Wired Networks |
| | Enhanced BAAL | No | Yes | Suitable to Wired Networks |
| | OMCT | No | Yes | Suitable to Wireless Networks |

In Static clustering approach, the multicast group is initially divided into several subgroups. Each subgroup shares a local session key managed by LC. Example: IOLUS [15] and DEP [5] belong to the categories, which are more scalable than centralized protocol.

Dynamic clustering approach aims to solve the "1 affect n" phenomenon. This approach starts a multicast session with centralized key management and divides the group dynamically. Example: AKMP [6], SAKM [16] belong to this approach and are dedicated to wired networks. Enhanced BAAL [17] and OMCT [7,8] proposes dynamic clustering scheme for multicast key distribution in adhoc networks.

OMCT [7,8] (*Optimized Multicast Cluster Tree*) is a dynamic clustering scheme for multicast key distribution dedicated to operate in ad hoc networks. This scheme optimizes energy consumption and latency for key delivery. Its main idea is to elect the local controllers of the created clusters [7,8]. OMCT needs the geographical location information of all group members in the construction of the key distribution tree.

Once the clusters are created within the multicast group, the new LC becomes responsible for the local key management and distribution to their local members, and also for the maintenance of the strongly correlated cluster property. The election of local controllers is done according to the localization and GPS (Global Positioning System) information of the group members, which does not reflect the true connectivity between nodes.

Optimized Multicast Cluster Tree with Multipoint Relays (OMCT with MPR) [18], whose main idea is to use information of Optimized Link State Routing Protocol (OLSR) to elect the local controllers of the created clusters. OMCT with MPRs assumes that routing control messages have been exchanged before the key distribution. It does not acknowledge the transmission and hence results in retransmission which consumes more energy.

Based on the literature reviewed, OMCT is the efficient dynamic clustering approach for secure multicast distribution in mobile adhoc networks. To enhance its efficiency, it is necessary to overcome the criteria, as OMCT needs geographical location information in the construction of key distribution tree by reflecting true connectivity between nodes.

## V. Discussions

In centralized protocols GKMP achieves an excellent result for storage at the members. However this result is achieved by providing no method for rekeying the group after a member has left, except re-creating the entire group which induces O(n) rekey message overhead where 'n' is the number of the remaining group members. Secure Lock achieves also excellent results for storage and communication overheads on both members and the key server. However, these results are achieved by increasing the computation overhead at the key server due to the Chinese Remainder calculations.

Distributed key agreement protocols do not rely on a group leader have an advantage over those with a group leader because, without a leader, all members are treated equally and if one or more members fail to complete the protocol, it will not affect the whole group. In the protocols with a group leader, a leader failure is fatal for creating the group key and the operation has to be restarted from scratch. The 1-affects-n phenomenon is not considered because in distributed protocols all the members are contributors in the creation of the group key and hence all of them should commit to the new key whenever a membership change occurs in the group.

In Decentralized protocols, protocols belong to the static clustering approaches are more scalable than centralized protocol. These protocols are dedicated to operate within wired networks.

Dynamic clustering approach aims to solve the "1 affect n" phenomenon. Dynamic clustering scheme are well suited for multicast key distribution in adhoc networks. OMCT (Optimized Multicast Cluster Tree) is a dynamic clustering scheme for multicast key distribution dedicated to operate in ad hoc networks. This scheme optimizes energy consumption and latency for key delivery.

## VI. Conclusion

Secure multicast communication is a significant requirement in emerging applications in adhoc environments like military or public emergency network applications. Membership dynamism is a major challenge in providing complete security in such networks. This dynamicity affects considerably the performance of the key management protocol. Most of the protocols suffer from 1-affects-n phenomenon.

This paper presents challenges, constraints and requirements for securing multicast key distribution for mobile ad hoc networks. It also presents taxonomy of key management protocols. This paper suggests OMCT (*Optimized Multicast Cluster Tree*) is a scalable scheme, which provides secure multicast communication in mobile adhoc network. This scheme is based on simple technique of clustering and key management approach. Thus this approach is scalable and efficient for dynamic multicast groups.

## AUTHORS PROFILE

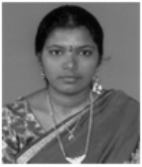

D. Suganya Devi received her B.Sc (Chemistry) and MCA from PSGR Krishnammal College for Women, Coimbatore in 1996 and 1999 respectively. And, she received her M.Phil degree in Computer Science in the year of 2003 from Manonmaniam Sundaranar University, Thirunelveli. She is pursuing her PhD at Avinashilingam University for Women. She is currently working as an Assistant Professor in the Department of computer Applications, SNR Sons College, Coimbatore. She has 10 years of teaching experience. She has presented 15 papers in various national, international conferences and journals. Her research interests Multicast Communication, MANET and Network Security.

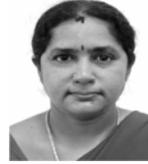

Dr. Padmavathi Ganapathi is the professor and head of Department of Computer Science, Avinashilingam University for Women, Coimbatore. She has 21 years of teaching experience and one year Industrial experience. Her areas of interest include Network security and Cryptography and real time communication. She has more than 60 publications at national and International level. She is a life member of many professional organizations like CSI, ISTE, AACE, WSEAS, ISCA, and UWA.